\documentclass{article}

\usepackage{arxiv}

\usepackage[utf8]{inputenc} 
\usepackage[T1]{fontenc}    
\usepackage{hyperref}       
\usepackage{url}            
\usepackage{booktabs}       
\usepackage{amsfonts}       
\usepackage{nicefrac}       
\usepackage{microtype}      
\usepackage{lipsum}		
\usepackage{graphicx}
\usepackage{doi}
\usepackage{amsmath}

\newcommand\chapter{\section}

\title{Benchmarking the performance of a self-custody
non-ledger-based obliviously managed digital
payment system}

\author{
    {William Macpherson}\\
	Department of Computer Science\\
	University College London\\
	\And
	{Geoff Goodell} \\
	Department of Computer Science\\
	University College London\\
 \\
}

\renewcommand{\shorttitle}{\textit{arXiv} Template}

\hypersetup{
pdftitle={A template for the arxiv style},
pdfsubject={q-bio.NC, q-bio.QM},
pdfauthor={David S.~Hippocampus, Elias D.~Striatum},
pdfkeywords={First keyword, Second keyword, More},
}

\begin{document}
\maketitle

\begin{abstract}
As global governments intensify efforts to operationalize retail central bank digital currencies (CBDCs), the imperative for architectures that preserve user privacy has never been more pronounced. This paper advances an existing retail CBDC framework developed at University College London. Utilizing the capabilities of the Comet research framework, our proposed design allows users to retain direct custody of their assets without the need for intermediary service providers, all while preserving transactional anonymity. The study unveils a novel technique to expedite the retrieval of Proof of Provenance, significantly accelerating the verification of transaction legitimacy through the refinement of Merkle Trie structures. In parallel, we introduce a streamlined Digital Ledger designed to offer fast, immutable, and decentralized transaction validation within a permissioned ecosystem. The ultimate objective of this research is to benchmark the performance of the legacy system formulated by the original Comet research team against the newly devised system elucidated in this paper. Our endeavour is to establish a foundational design for a scalable national infrastructure proficient in seamlessly processing thousands of transactions in real-time, without compromising consumer privacy or data integrity.
\end{abstract}

\section{Introduction}
\pagenumbering{arabic}
The decline of physical cash as a payment method presents an increasingly urgent issue for contemporary consumers. In a world where digital card payments are becoming the norm, both consumers and merchants are relinquishing the inherent advantages that come with cash transactions. Cash payments, for instance, offer complete anonymity for the consumer and are often fee-free for merchants. However, digital card payments effectively strip these parties of such benefits: consumers' personal data is frequently shared among multiple intermediaries, while merchants face high transaction fees to facilitate these payments.

In a landscape increasingly dominated by large corporations, the modern consumer's quest for convenience has led to a digital payments ecosystem where companies profit from both user data and transaction fees. This paradigm shift has underscored the importance of speed in transactions, as well as the consumer's desire for privacy, as evident in the growing influence of blockchain technology and cryptocurrencies.

Building on this contextual backdrop, this paper aims to extend an existing digital payments framework developed by the Comet Project team~\cite{goodell2022scalable,comet2023}. Specifically, we benchmark the transactional speed of the Comet system, aiming to deliver high-speed transactions in an environment designed with privacy as a core principle. We also introduce a novel approach for improving transactional speed by altering the method in which Merkle Tries are built to cumulatively add transactions, this altercation theoretically improves the time it takes to validate a transaction's legitimacy.

In utilizing a quantitative methodology, we strive to find an architectural equilibrium where computational limitations are acceptable for a scalable national payment ecosystem, and transaction speed can compete with the latest retail payment systems. To achieve this, we employ a targeted approach centred around parameter tuning. Through systematic adjustments to key system variables, we aim to identify the optimal settings that balance high transactional speed with computational efficiency. This fine-tuning is instrumental for assessing the system's scalability and for envisioning its real-world applicability, particularly in the context of a central bank digital currency. Our work aspires to bridge the gap between user privacy and transactional efficiency, laying the groundwork for a more equitable digital payments landscape.

\subsection{Background on Central Bank Digital Currencies}
In recent years, central bank digital currencies (CBDCs) have gained prominence, driven by the digitization of economies and the increasing influence of cryptocurrencies \cite{boe2023,ecb2023}. The Bank of England (BoE) proposes a comprehensive blueprint for a digital pound, emphasizing offline payments, user privacy, interoperability, and a robust legal framework \cite{houseoflords2022cbdc,boe2023}. Similarly, the European Commission advocates for a digital Euro with a focus on regulatory clarity, integration with existing financial systems, and privacy compliance using Distributed Ledger Technology (DLT) \cite{ecb2023}. A comparative analysis reveals shared priorities in regulatory frameworks and consumer protection, with the BoE providing more detailed insights into operating models and programmability \cite{boe2023,ecb2023}. Recent studies underscore the need for stringent regulatory frameworks, consumer protection, and seamless integration with existing payment infrastructures \cite{chaum2021cbdc}. In conclusion, both the BoE and the European Commission aim to establish secure, efficient, and interoperable digital currency infrastructures, aligning with modern technological advancements and regulatory norms \cite{boe2023,ecb2023}. However, debates surrounding the digital pound raise concerns about privacy, surveillance, security, and economic repercussions, emphasizing the need for a meticulous assessment of the complex landscape surrounding the introduction of a national digital currency \cite{boe2020cbdc,houseoflords2022cbdc,abelson2015keys,goodell2022scalable,goodell2023}.

\section{The Comet Project}
\subsection{Introduction}
The Comet Project~\cite{comet2023} represents a new approach to electronic retail payments utilizing central bank digital currency (CBDC) \cite{boe2020cbdc}. By adeptly navigating the delicate balance between rigorous regulatory supervision and essential consumer privileges, such as privacy and autonomy, Comet paves the way for a paradigm shift in the domain of retail CBDCs.  Comet serves as the foundation for the research embodied in this paper.
At its essence, Comet champions a ``privacy by design'' philosophy, paving the way for the evolution of retail CBDC. By harnessing the unique potential of the USO asset model \cite{goodell2022scalable}, it facilitates a platform where consumers enjoy complete anonymity coupled with unyielding sovereignty over their assets.
\subsection{Architectural Insight}
The Comet architecture encompasses six pivotal systems: the Consumer System, MSB Service, PoS Service, Relay Service, Minting Service, and the CB Service. Each system performs a distinct, yet interlinked, function within the overarching ecosystem, we focus exclusively on the Relay Services functionality within this paper

\begin{figure}
    \centering
    \includegraphics[width=1\linewidth]{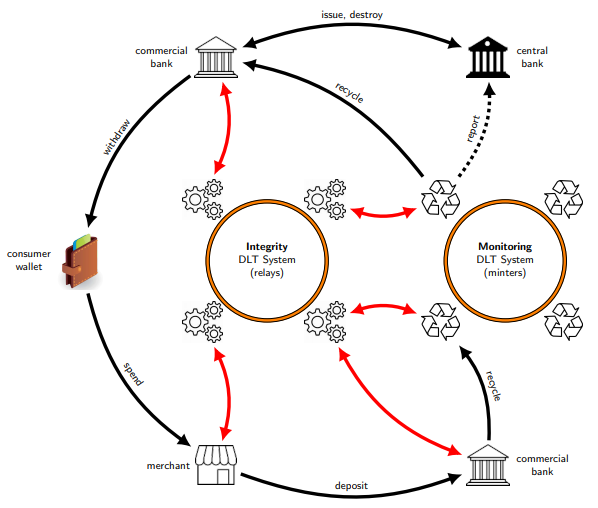}
    \caption{Comet System Design}
    \label{fig:cometdesign}
\end{figure}

\subsection{Relay Service}
To fully grasp the enhancements we implemented in the Relay Service for swifter creation of Proofs of Provenance (PoP) we must present the conventional structure and function of the Relay Service. It is instrumental in facilitating PoP's and transmitting Merkle Roots to the DLT system.

A Merkle Root is derived from a Merkle Trie, a binary tree data structure that houses key-value pairs. In this configuration, the key aids in pinpointing the specific location where the value resides within the trie. The generation of a Merkle root entails a hierarchical hashing process initiated at the leaf nodes, progressing upward by pairing and hashing adjacent nodes until the root node is reached. This root node, therefore, embodies a unique hash representing all the key-value pairs stored in the trie, ensuring a comprehensive data representation.

The Relay Service operates in distinct intervals, each characterized by a defined timeframe termed a "cycle." Within each cycle, it aggregates key-value (KV) pairs relayed from various point-of-sale devices. This aggregation forms the basis of the Cycle Trie, the Merkle Trie data structure that essentially amasses all transaction details within that period. Given the abstract nature of a Merkle Root, it is devoid of tangible transactional data, thus safeguarding transaction confidentiality and security.

Post cycle completion, the Relay Service undertakes the construction of the Cycle Trie, integrating all received KV pairs before archiving it. Simultaneously, a separate pathway is initiated to transfer these Merkle Roots to the DLT system. Here, a consensus on the roots is achieved, facilitating their incorporation into the ledger, thereby enhancing transparency and security.

These Cycle Tries are central to the architectural blueprint of proofs of provenance, offering a reliable mechanism to ascertain asset legitimacy and ownership. Leveraging proofs, asset holders can confidently verify the authenticity of their holdings, fostering a secure transactional environment. Detailed insights into proofs are described in later sections.

\label{Chap3}
\section{Blockchain Technology and USO Assets}
Blockchain technology, inherently a form of Distributed Ledger Technology (DLT), emerged as a ground-breaking advancement in the digital realm, spearheaded by the inception of Bitcoin in 2009 followed by Ethereum in 2015 \cite{nakamoto2008bitcoin,wood2014ethereum}. At its core, DLT thrives on decentralization, employing multiple autonomous nodes to maintain and update individual instances of a ledger. These nodes synchronize periodically to establish consensus and safeguard data integrity throughout the network \cite{zheng2018blockchain}.

A blockchain materializes as a series of blocks linked sequentially, wherein each block contains a batch of transactions. Commencing with the 'Genesis Block', each succeeding block is securely and irrevocably linked to its predecessor through cryptographic hashes, establishing an immutable ledger. This section sets out the foundational concepts of blockchain technology, with an emphasis on the distinctive attributes of Ethereum and Bitcoin blockchains, and explores the implications of permissioned and permissionless blockchain designs.

\subsection{USO Assets}
The use of unforgeable, stateful, oblivious (USO) assets, exemplified by the TODA protocol~\cite{TODAIntroduction}, emerges as a new paradigm in the digital landscape, presenting a solution akin to presenting the digital equivalent of a piece of paper, a concept that is pivotal in understanding the foundation of this protocol. In this regard, the protocol operates based on a principle that underscores transferability as a vital element, facilitating transactions that are secure and trustworthy.

As we can see from the description of the TODA protocol~\cite{TODATechPrimer}, USO assets are intended to dismiss the intrinsically limiting aspects seen in replicated ledgers, standing tall as a protocol that enhances both security and efficiency. The intrinsic architecture of TODA allows it to function devoid of a central authority, thereby relinquishing control from a central entity and supporting a space that is free of overarching control and surveillance of the flow of assets, a feature that is quintessential in ensuring the autonomy of transactions.

One of the standout features of the USO asset paradigm is its emphasis on maintaining integrity. This is facilitated through a system where the objects can maintain integrity through one system while having their state fully managed by another. This design not only ensures the secure management of objects but also fosters a ground for maintaining unequivocal integrity. Furthermore, the protocol accentuates the importance of transferability, a feature that is crucial in establishing trust in digital transactions. It eradicates the necessity for cumulative trust that is generally seen in a chain of transfers, thereby presenting a straightforward and secure mechanism for transfers.

USO assets stand as a revolutionary architecture in the digital space, bringing forth a plethora of features that are grounded in ensuring security, transferability, and integrity, thereby promising a landscape that is not only secure but also autonomous.  The USO asset paradigm has substantial potential for the implementation of Central Bank Digital Currencies, offering a secure and efficient platform that could potentially revolutionize payment ecosystems globally by providing a robust infrastructure that upholds the principles of security and integrity, which are cardinal to the future development of CBDC.

\subsection{Merkle Trie}
The Merkle Trie, a variant of the Trie data structure, named for its retrieval properties, incorporates cryptographic hashing, leveraging the principles of Merkle trees to facilitate verifiable proofs regarding the existence or non-existence of certain values within associative arrays, where keys are predominantly strings. This integration bestows unique properties upon Merkle Tries, distinguishing them fundamentally from regular trees and arrays.

\subsubsection{Structure and Functionality}
A Trie, fundamentally a tree-like data structure, delineates each path from the root to a leaf as a key, storing the corresponding value at the leaf. The Merkle Trie extends this structure by associating a cryptographic hash with every node, enhancing security and verifiability. The structure can be detailed as:
\begin{itemize}
    \item \textbf{Leaf Nodes}: These nodes house the actual values, accompanied by a hash of the value.
    \item \textbf{Internal Nodes}: These nodes, devoid of direct data storage, contain hashes derived from a combination of their child nodes' hashes, facilitating a cryptographic linkage and integrity verification from leaf to root.
\end{itemize}

\subsubsection{Verifiable Proofs}
The Merkle Trie's quintessence is its ability to generate verifiable proofs pertaining to the existence or non-existence of a value, leveraging the cryptographic hashes stored in its nodes. This feature enables efficient and secure queries; a user can verify a value's status by accessing a subset of hashes, referred to as a "trimmed trie," rather than the entire Trie, thereby attesting to the value's presence and its untampered state~\cite{TODATechPrimer}.

\subsubsection{Application}
In the realm of distributed systems, especially blockchain technologies that prioritize data integrity and verifiability, Merkle Tries hold a pivotal role. While Merkle trees facilitate a reliable avenue for consolidating and verifying transactions, certain blockchains like Ethereum necessitate a tool to encapsulate the system's comprehensive state, including account balances and contract codes. Here, Merkle Tries step in, offering a compact and secure representation, an upgrade from Merkle trees, thereby ensuring an efficient and secure digital asset management system \cite{buterin2014ethereum}.

\subsection{Proof with Cycle Tries}
Proofs of Provence (PoP) in point-of-sale (PoS) systems are integral to affirming the non-existence of a transaction, a null proof, validating the asset's ownership by the merchant and negating double spending issues prevalent in traditional blockchains and cryptocurrencies \cite{nakamoto2008bitcoin}. The speed of retrieving a PoP hinges on the number of cycles and the time frame in question.

The process involves the PoS device requesting the relay service to verify the transaction from its inception to the current cycle, through a rigorous check of every root. This system operates based on the computational complexity given by
\[
O(q\times\log\left(\frac{n}{q}\right))\ =\ q\times O(\log(n))-q\times O(\log(q))
\]
where \(n\) represents the number of transactions in the period, and \(q\) stands for the number of cycles per period, dictating the depth of the Merkle Trie for a single cycle and consequently the number of hash transactions required for proof construction over all \(q\) cycles.

This mechanism, despite its efficiency in negating fraudulent transactions, poses a challenge: a surge in the cycles time can potentially delay confirmation, a caveat requiring a delicate balance to maintain system efficiency while ensuring transaction security.

From the moment the card is swiped or inserted to the time it receives approval from the issuing bank, a series of critical evaluations ensue, corroborated by necessary security checks. Yet, despite technological advancements making this a seemingly swift process, the settlement can often take up to three days, resulting in a deferred settlement. It indeed is a test of trust where the consumer implicitly believes in the issuing bank's efficiency, and the merchant trusts the acquiring bank to confirm payments swiftly and transfer funds later, albeit at the cost of potential cash flow challenges~\cite{federal_reserve_2019}.

\section{Cumulative Merkle Trie Design}
To enhance the efficiency of Proof of Provenance (PoP) retrieval, we propose a novel approach: the Cumulative Merkle Trie design, a possible improvement to the traditional PoP retrieval incorporated into the TODA Architecture \cite{TODATechPrimer} and implemented in the traditional Relay Service \cite{goodell2022scalable}. This legacy system accumulates Key-Value (KV) pairs from PoS devices every cycle, constructing a Cycle Trie at the end of said cycle. Despite ensuring transaction security and anonymity, it renders the proof retrieval process, particularly the null proof spanning multiple cycles, considerably lengthy.

Our Cumulative Merkle Trie design is aimed to overcome this bottleneck, aggregating transactions in a continually updated Trie, with the Merkle Root of each period encapsulating all previous transactions since the inception of the Trie. This facilitates a streamlined PoP retrieval, where the PoS device merely validates the transaction's existence in a specific period, contrasting it against its non-existence in the preceding cycle, a process implicitly affirmed if the PoP retrieval occurs within the same cycle of the transaction's addition to the Trie.

This nuanced modification offers prominent advantages:
\begin{enumerate}
    \item \textbf{Speed and Efficiency}: It substantially reduces the PoP retrieval time, expediting business operations and bolstering trust among users.
    \item \textbf{Reduced Computational Overhead}: It mitigates the necessity to traverse numerous roots for PoP computation, thereby conserving computational resources, a vital asset in high-transaction environments.
\end{enumerate}
\begin{figure}
    \centering
    \includegraphics[width=1\linewidth]{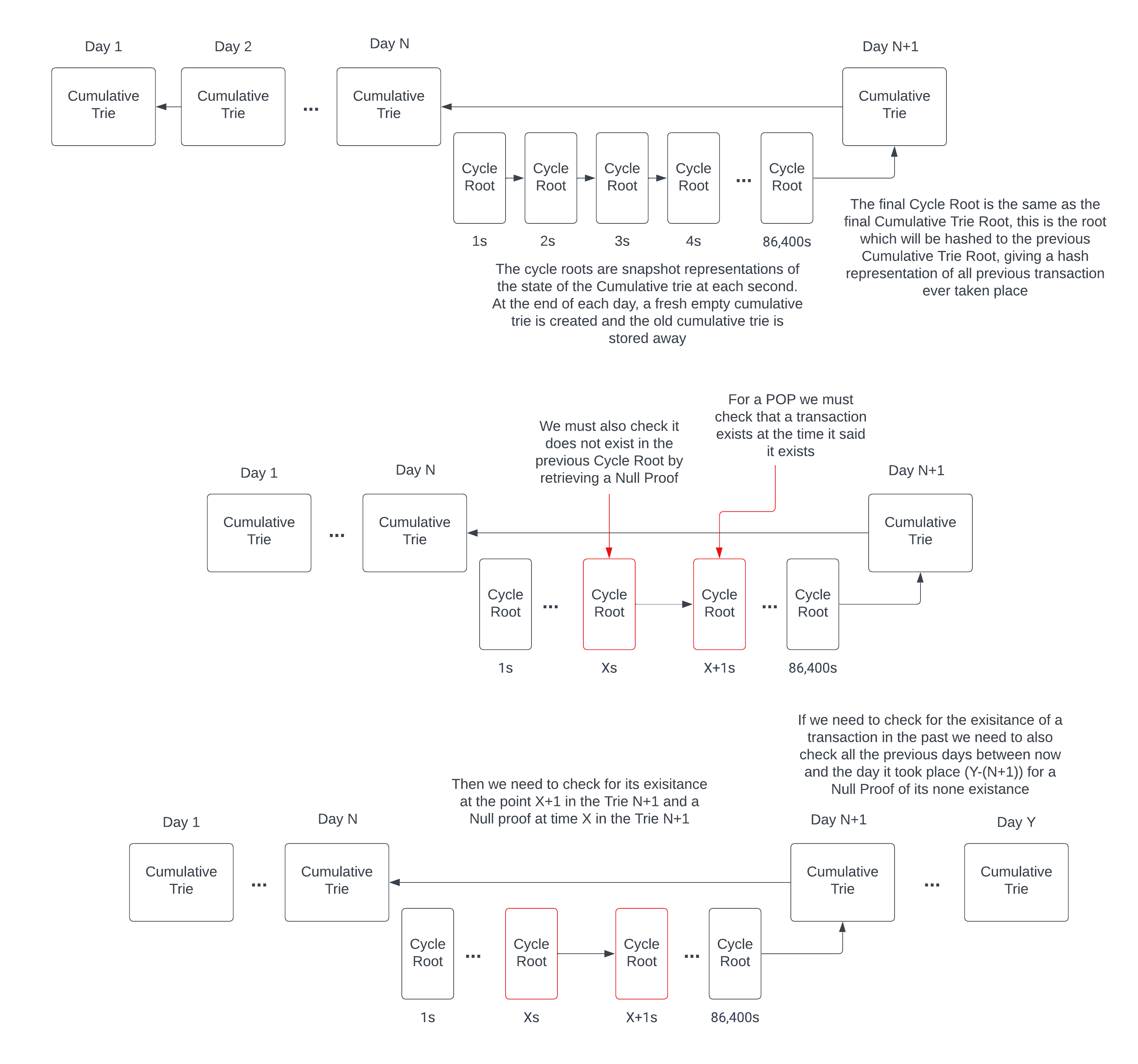}
    \caption{Cumulative Merkle Trie Proof Retrieval, Cycle Time is 1s, Cumulative Refresh Rate 1 day}
    \label{fig:cumTrieproof}
\end{figure}
Intrinsically, this design entails a deeper traversal into the Merkle Trie, as the transaction volume, denoted as \(n\), linearly escalates over cycles. However, this depth increment follows a logarithmic pattern, mathematically expressed as \(O(\log(n))\), depicting that the computational exertion for proof retrievals augments logarithmically as \(n\) linearly grows, a trait ensuring the system's high efficiency and fostering greater scalability without compromising security.

To ensure manageable computational overhead, we predicate on an assumption: the relay service, upon receiving a transaction inclusion request from a merchant, reciprocates with a confirmation alongside the transaction's proof in the cumulative Merkle Trie. This approach obviates the need for multiple Trie replicas, averting an exponential surge in storage requisites calculated as \(x^n\), with \(n\) representing the cycle count and \(x\) the constant transaction number in each cycle.

Despite its merits, the design imposes considerable storage demands due to the continual expansion of the Merkle Trie, posing a challenge for entities with limited computational resources. However, for substantial entities like Mastercard or Visa for which we envision this relay service to be maintained by, the expansive storage requisite is manageable, justifying the trade-off between augmented efficiency and heightened storage needs, thereby theoretically championing a viable and rewarding solution.
\section{Blockchain Redesign for Enhanced Efficiency and Focused Functionality}
In the rapidly evolving landscape of distributed ledger technologies (DLTs), there is a pressing need to devise systems that surpass existing configurations in terms of efficiency and functionality. Despite the availability of a multitude of open-source blockchains, a critical evaluation of our specific requirements drove us to architect a DLT system that prioritizes the fundamental ledger functionality, facilitating a decentralised and immutable record to enhance resilience and trustworthiness. 

Our DLT initiative departs from the conventional permissionless setting, opting instead for a permissioned environment limited to regulated entities. This curated approach not only enhances security through mitigating Sybil attacks but also allows for the omission of computationally intensive consensus mechanisms like Proof of Work and Proof of Stake fostering a more efficient system.

Rooted in principles of decentralization and immutable record-keeping, our DLT system abandons elements traditionally seen as foundational in blockchain technology. The exclusion of superfluous features such as smart contracts and token management results in a streamlined ledger with an agile, high-performing, and lightweight operational framework, without compromising on core efficiency.
\subsection{The Raft Consensus Algorithm}
Our choice of the Raft consensus algorithm, conceptualized by Ongaro and Ousterhout~\cite{ongaro2014search}, represents a strategic move towards a simplified yet effective consensus model. The Raft algorithm incorporates key mechanisms, including leader election through terms, promoting democratic leadership via majority votes. It ensures a uniform log state, emphasizes stringent safety rules for servers during leadership transitions, employs a two-phased approach for membership changes ensuring system stability, and centralizes read and write requests management to leaders, ensuring efficient and reliable system operations.
\subsection{Performance of Raft in Private Blockchains \cite{8666147}}
Private blockchains demand a consensus algorithm that is both reliable and efficient. The performance analysis of the Raft algorithm in private blockchains reveals it as a suitable candidate, demonstrating admirable throughput and latency characteristics compared to other consensus algorithms like PoW and PoS,

The Raft consensus algorithm showcases higher throughput and lower latency, enhancing the performance metrics in private blockchains. Its architecture facilitates faster block generation and validation processes, making it a preferred choice for real-time applications.

Raft exhibits a balanced resource utilization strategy, making efficient use of system resources. It operates with lower computational power, thereby reducing operational costs and fostering a sustainable blockchain ecosystem.

The Raft algorithm stands tall in ensuring safety and reliability in private blockchains. Its fault tolerance capabilities and robust safety features make it a reliable choice for maintaining the integrity of private blockchain networks.
\subsection{Choosing the Raft Algorithm}
Based on the performance analysis \cite{8666147}, the Raft algorithm emerges as a favourable choice for this project. Its high throughput and low latency characteristics, coupled with efficient resource utilization and robust safety features, make it a promising candidate for private blockchain implementations. Furthermore, its emphasis on simplicity makes it a suitable fit for an off-the-shelf solution to our DLT needs.

The algorithm offers clear advantages, including streamlined leadership election, efficient log replication, and strong safety guarantees, positioning it as a reliable choice for the project. Its performance metrics in private blockchains further reinforce its suitability for the project, promising an efficient and reliable system \cite{8666147}.

Despite its advantages, the Raft algorithm comes with its set of challenges, including potential vulnerabilities in dynamic environments. However, with appropriate mitigative strategies, these challenges can be effectively addressed, paving the way for a successful implementation for our DLT needs \cite{8666147}.
\section{Technical Methodology for DLT System Implementation}
In the development of our Distributed Ledger Technology (DLT) system, we prioritized simplicity and speed in achieving consensus. To this end, we utilized the Java programming language and the Spring framework to facilitate continuous deployment. Our lightweight container, created through this approach, can be efficiently deployed across various computers, representing different nodes. This design is split into two significant components: the Raft consensus algorithm and node architecture.
\subsection{Raft Consensus Algorithm\cite{8666147, ongaro2014search}}
Maintaining consistency across numerous nodes is a pivotal challenge in distributed systems. To address this, we employed the Raft consensus algorithm, which ensures reliable operations and data consistency across multiple servers, even amidst failures. A cornerstone of this algorithms implementation is the atomic register, a data structure that promotes high reliability and consistency through atomic read and write operations.
\subsubsection{Atomic Register}
Central to the operation of our DLT system is the atomic register, which supports only two basic operations: read and write. These operations are carried out atomically, meaning that they are indivisible and instantaneous, ensuring a high degree of consistency and reliability in the system. Utilizing an atomic register mitigates errors arising from concurrent operations, thereby enhancing the robustness of the system.

Returning to our design, the node structure harmonizes with the atomic register and Raft consensus algorithm to secure system stability. The consensus process initiates with a request from the Relay Service to a node in the Raft group. Should the selected node not be the leader, it redirects the request to the group leader, who circulates it amongst all group nodes to obtain consensus. This approach ensures that a majority of nodes concurs on the data state, a prerequisite maintained by the Raft algorithm for sustaining consistency across the distributed ledger.

Upon reaching consensus, the leader node and follower nodes commit the Merkle Root contained in the original request to the atomic their respective registers. Leveraging the MicroRaft implementation in Java \cite{microraft2021} facilitates the realization of a compact, efficient, and reliable distributed system through a robust library. This strategy allows for the retention of the Java programming environment, familiar and predominant in our system architecture, with the atomic register underpinning the node operations with steadfast consistency and reliability.
\subsection{Node}
In our DLT system's architecture, the node is a central component, meticulously constructed to foster consistency and replication. Capitalizing on Java's versatility and the robust Spring framework, the node promises streamlined deployment and operational stability. A defining characteristic is the integration of environmental variables directly into the codebase, allowing users to tailor pivotal parameters during the initialization phase, encompassing port assignments, MongoDB backend configurations, and crucial Raft group connection details.

At its core, the node operates through a listener mechanism, attentively monitoring a pre-defined global variable for any changes in state. This mechanism is primed to activate upon the completion of an atomic register set operation, intrinsically tied to the modification of the monitored global variable. Subsequently, the node launches a retrieval operation to extract the recently updated Merkle Root from the atomic register.

The new Merkle Root's storage as a local variable initiates a series of operations required to create a new block, which includes gathering vital data elements such as the previous block's hash, a newly calculated index, the precise creation timestamp, and the Merkle Root itself. The SHA-256 algorithm is then deployed to hash the assembled data, generating a unique block ID and inherently forming a linked chain, a fundamental characteristic of blockchain technology.

Once formed, the new block is stored in the node's individual MongoDB database, a step replicated across all nodes in the Raft group to uphold decentralization. This synchronous process embodies the essence of blockchain, where transparency, redundancy, and immutability are central features.

Furthermore, the node undertakes a secondary role as a query endpoint for the relay service and authorized entities, facilitating transaction verification through Merkle Root checks on the blockchain. This function is essential in maintaining the system's integrity by leveraging the cumulative nature of the Trie structure to validate transactions against historical data recorded in preceding cycles.
\begin{figure}
    \centering
    \includegraphics[width=1\linewidth]{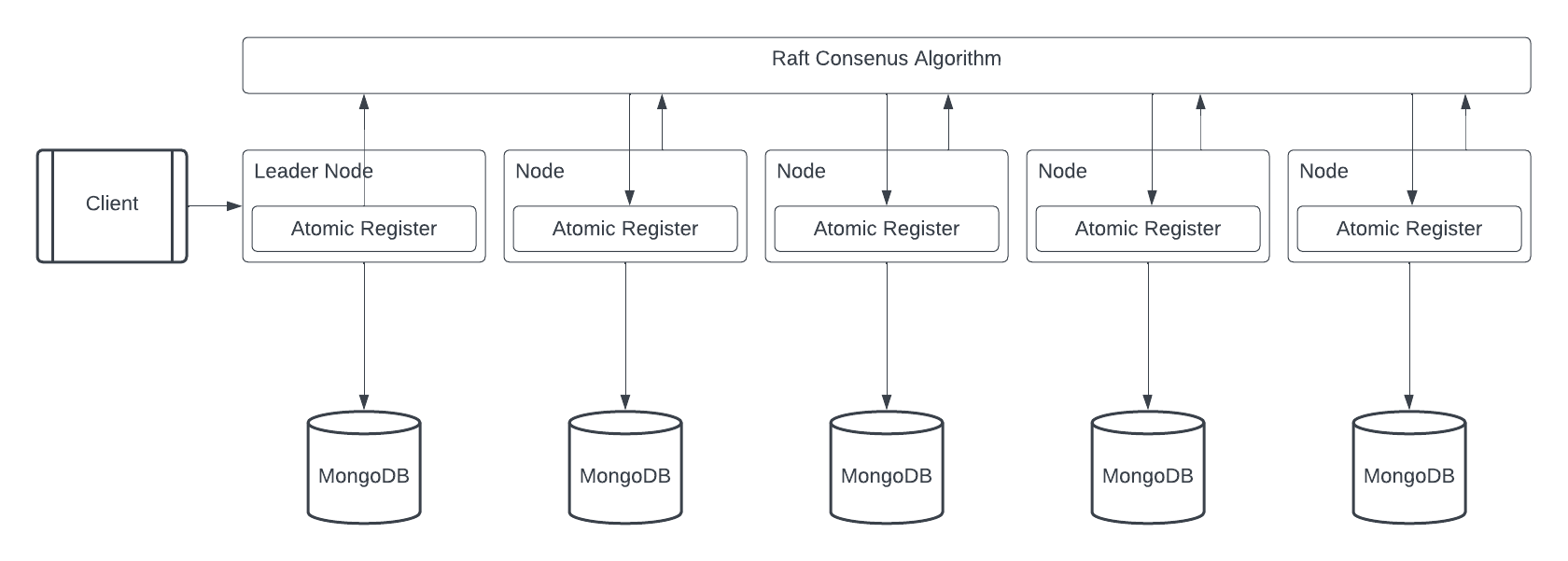}
    \caption{Raft System Implementation}
    \label{fig:raftimplementation}
\end{figure}
\subsection{Differentiating Block Formation from Ethereum and Bitcoin}
Our DLT system shares foundational principles with blockchain technologies like Ethereum and Bitcoin; however, it diverges in terms of block formation and structure.

Unlike the traditional frameworks of Ethereum and Bitcoin, where blocks primarily encapsulate a series of transactions to maintain a detailed history of network activities, our system focuses on the Merkle Root. This approach bypasses the need for storing individual transactions in each block, opting instead for a cryptographic summary derived from the Merkle Tree structure. This strategy not only simplifies the validation process but also substantially reduces storage requirements, thereby crafting a leaner and more efficient database.

In summary, our DLT system marries the security and integrity synonymous with blockchain technology with proposed enhanced efficiency and a reduction in computational demands.
\section{Relay Service Implementation}
\subsubsection{Framework and Infrastructure}
The relay service is also architected in Java leveraging the Spring framework to facilitate continuous deployment. It's designed to listen on a pre-configured port. For our backend database, a MongoDB instance is containerized using Docker, which stores each cycle root as well as the entire cumulative Merkle Trie at the end of a designated period.
\subsubsection{Data Collection and Processing}
The application accumulates key-value pairs in a synchronized fashion that aligns with the addition to the cumulative Merkle Trie. The service's transaction processing capacity is intrinsically tied to the speed of additions to the Merkle Trie. If additions to the Trie lag, the real-time queue of key-value pairs could swell, potentially overwhelming the Trie builder.  However, it's crucial to recognize the variable nature of retail transactions across the UK, transactional volume is not uniform throughout the day. For instance, transactional peaks often occur around 13:00, while 1:00 sees a lull. This fluctuation can provide some computational buffer, allowing the system to manage transactional surges during peak times and catch up during quieter intervals.

As referenced in our Cumulative Merkle Trie design, we are making the assumption that proofs are returned to the sender within the same cycle to which the key-value pairs were added to the Cumulative Trie. Due to this assumption, we must take into account the processing time it takes to form each of these proofs with the aim of calculating the latency of the relay service. We can then mathematically represent the Relay's latency as \(
L = R_{ctp} + R_{pr}
\), we will allude to this within our next section focusing on theoretical analysis.

\section{Influencing Criteria and Theoretical Analysis}
To critically understand and benchmark the performance of our Cumulative Merkle Trie design, we must identify and analyze the fundamental parameters that influence its efficiency and overall performance. In this section, we delve deep into each parameter, establishing mathematical relationships between them with the ultimate goal of enhancing the Proof of Provenance (PoP) retrieval process. 

\subsection{Incoming Number of Transactions ($n$)}
This parameter represents the number of transactions that the relay service accumulates over a single cycle. The value of $n$ can theoretically be any non-negative integer. Its impact on the system can be analyzed as:
\[
D = \log_2(n)
\]
Where:
\begin{itemize}
    \item $D$: Depth of the Merkle Trie
    \item $n$: Incoming number of transactions
\end{itemize}
\subsection{Total Number of Transactions ($N$)}
Representing the sum of incoming transactions over all previous cycles within a period plus the incoming transactions in the current cycle, it's given by:
\[
N = n + \sum_{i=1}^{m-1} n_i
\]
And the depth of the Merkle Trie in terms of $N$ would be:
\[
D = \log_2(N)
\]
Where:
\begin{itemize}
    \item $N$: Total number of transactions
    \item $n_i$: Incoming transactions in the i-th cycle
    \item $m$: Number of cycles
\end{itemize}
\subsection{Depth of the Merkle Trie ($D$)}
The depth of the Merkle Trie is intrinsically linked to the number of transactions it holds, increasing logarithmically as the total number of transactions ($N$) increases. It is given by the formula mentioned above. A deeper Trie generally implies a more complex structure, affecting other parameters such as the retrieval processing rate and cumulative Trie processing rate.

\subsection{Cycle Time ($T_c$)}
Cycle time is the duration of a single cycle, during which transactions are accumulated. This parameter could influence the number of transactions accumulated ($n$) and thereby the depth of the Merkle Trie ($D$). It can be represented as:
\[
n = \lambda \cdot T_c
\]
Where:
\begin{itemize}
    \item $\lambda$: Average rate of incoming transactions
    \item $T_c$: Cycle time
\end{itemize}

\subsection{Period Time ($T_p$)}
Period time is the duration before a new cumulative Trie is created, expressed in terms of the number of cycles. It influences the overall processing and retrieval rates, and can be represented mathematically as:
\[
T_p = m \cdot T_c
\]
Further simplified: 
\[
T_p = m \cdot \frac{\lambda}{n} 
\]
Where:
\begin{itemize}
    \item $m$: Number of cycles
    \item $T_p$: Period Time
    \item $\lambda$: Average rate of incoming transactions
    \item $n$: Number of transactions accumulated in a cycle
\end{itemize}

\subsection{Cumulative Trie Processing Rate ($R_{ctp}$)}
This is the rate at which transactions are added to the existing Merkle Trie. It is expected to vary with the number of the transactions stored within the Merkle Trie ($N$) due to increased rehashing efforts as the Trie becomes larger. It can be modeled as:
\[
R_{ctp} = f(N)
\]
Where:
\begin{itemize}
    \item $f(N)$: A function representing the processing rate in terms total number of KV pairs stored
\end{itemize}

\subsection{Proof of Provenance Retrieval Processing Rate ($R_{pr}$)}
This parameter indicates the rate at which proofs can be retrieved from the stored Merkle Trie. Similar to the cumulative Trie processing rate, it is expected to vary with the number of KV pairs stored ($N$). It can be represented as:
\[
R_{pr} = g(N)
\]
Where:
\begin{itemize}
    \item $g(N)$: A function representing the retrieval rate in terms total number of KV pairs stored
\end{itemize}

\subsection{Relay Latency ($L$)}

Relay latency is the summation of the cumulative Trie processing rate and the proof of provenance retrieval processing rate, representing the overall computational delay in the system. It can be expressed as:
\[
L = R_{ctp} + R_{pr}
\]
Where:
\begin{itemize}
    \item $L$: Relay latency
    \item $R_{ctp}$: Cumulative Trie processing rate
    \item $R_{pr}$: Proof of Provenance retrieval processing rate
\end{itemize}

By identifying and mathematically representing the relationships between the different parameters influencing the Cumulative Merkle Trie system, we lay a theoretical foundation to guide empirical testing. Understanding these relationships will be pivotal in tuning the system to achieve optimal performance, with a special focus on reducing the Proof of Provenance (PoP) retrieval time. Our future work will focus on determining the functions $f(N)$ and $g(N)$ through experimental analyses to foster a more robust understanding of the system's dynamics. We shall also go one step further in attempting to identify the optimal parameters needed for an efficient system.

\section{Legacy Relay Service Benchmarking}
Our work to understand the relationships between parameters does not stop here. We must also understand the impact that our legacy archiecture had on these parameters to critically understand if our novel system provides a true speed up in PoP retrieval. Although many parameters mathematical relationships can be inherited from our novel solution there are some key differences within our Proof of Provenance Retrieval Processing Rate ($R_{pr}$), most notably we cannot express it in terms of the number of KV pairs stored within the Merkle Trie, \(N\). 

In the legacy system, the \(R_{pr}\) is essentially dictated by the number of cycles intervening between the inception of the asset and the transaction's assimilation into the current cycle. This necessitates a mathematical representation that encompasses the cycle difference parameter (\(\Delta C\)), defined as:

\[
\Delta C = C_{\text{current}} - C_{\text{inception}}
\]

Where:
\begin{itemize}
    \item \(C_{\text{current}}\): The cycle number representing the current cycle
    \item \(C_{\text{inception}}\): The cycle number during which the asset was conceived
\end{itemize}

Subsequently, the \(R_{pr}\) in the legacy system can be construed as a function of \(\Delta C\), expressed as:

\[
R_{pr}^{\text{legacy}} = h(\Delta C)
\]

Where:
\begin{itemize}
    \item \(h(\Delta C)\): A function representing the retrieval processing rate as a function of the cycle difference
\end{itemize}

This rendition of \(R_{pr}\) underscores a critical deviation from our novel system, incurring what we predict is a non-linear relationship. 

\subsection{Conclusion}
By dissecting the critical parameters and establishing mathematical relationships in both the novel and legacy systems, we have paved a foundational pathway for empirical testing. The subsequent phase involves meticulous benchmarking based on these theoretical underpinnings, steering towards an empirical validation of the novel system's superiority in PoP retrieval times. It is envisaged that this analytical journey will culminate in establishing a system that stands tall on parameters of efficiency and speed, thereby improving PoP retrieval processes.

\chapter{Results and Analysis} \label{Chap5}
In this chapter, we discuss the process undertaken to quantify the performance metrics of the relay service in terms of transaction processing speed ($R_{ctp}$) and the proof of provenance retrieval processing rate ($R_{pr}$). We employ two distinct methodologies to assess both the novel and the legacy systems. We subsequently show the results of this methodology and analyse the results.

\section{Transaction Processing Speed ($R_{ctp}$)}
\subsection{Novel System Evalutation}
For the experimentation, a Java test harness was developed to send simultaneous update requests to the relay service. These requests contained the key-value (KV) pairs to be incorporated into the cumulative Merkle Trie. A successful response from the relay service offers an assurance of the successful addition of KV pairs to the Merkle Trie. 

To facilitate a focused analysis, the cycle time was adjusted to a sufficiently large value to prevent overlap with successive cycles, thereby averting potential overflows. This setup was engaged in four separate runs with varying numbers of incoming transactions $n$, (25, 50, 75, 100), enabling the baseline determination of the service latency grounded on the total number of KV pairs within the Trie, aligning with the projections made in the preliminary methodology. 

\subsection{Novel $R_{ctp}$ Results and Analysis}
\begin{figure}
    \centering
    \includegraphics[width=1\linewidth]{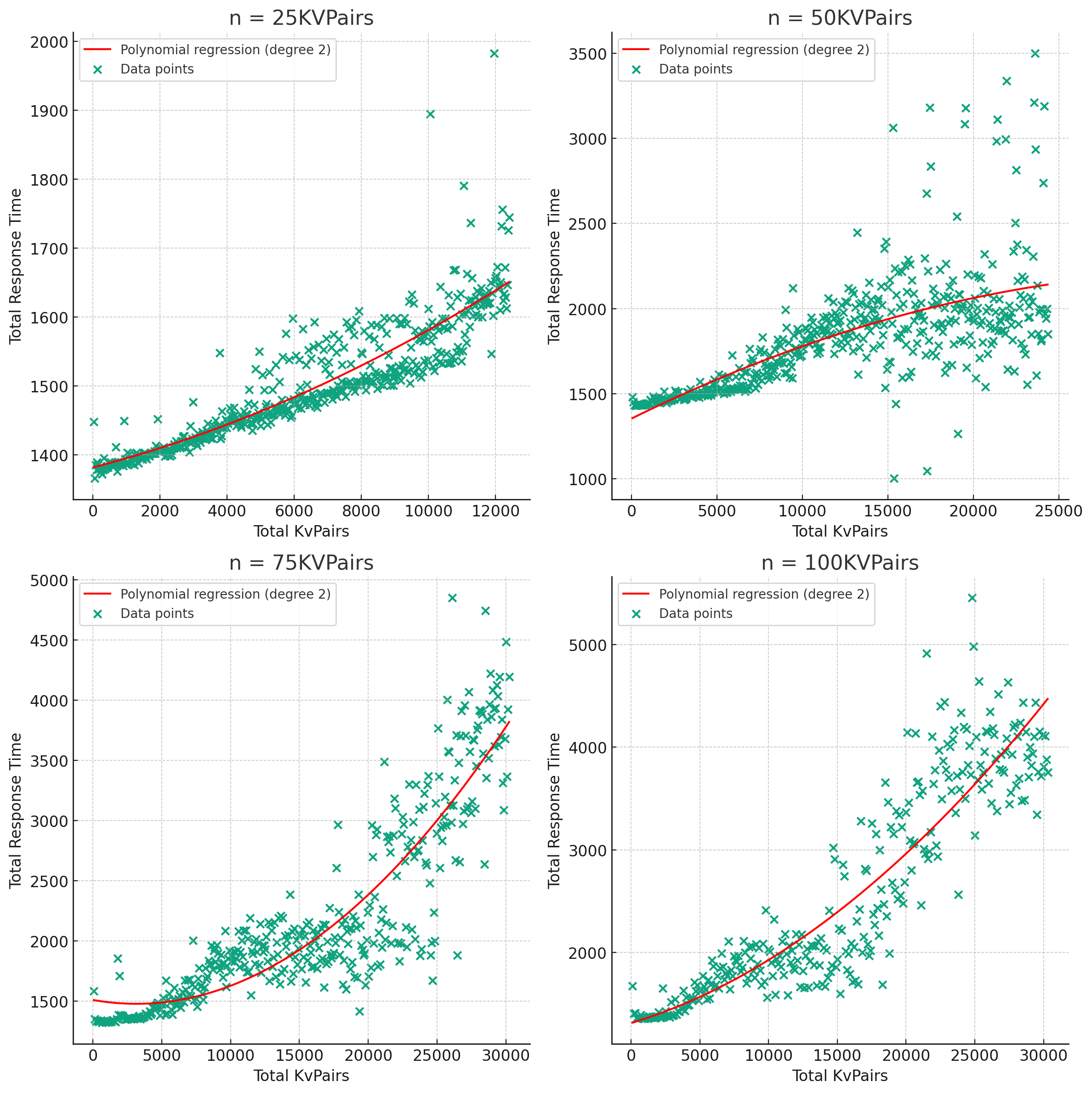}
    \caption{Total Response Time vs Total KV Pairs for 4 different values of n}
    \label{fig:4datasets}
\end{figure}
\begin{figure}
    \centering
    \includegraphics[width=0.75\linewidth]{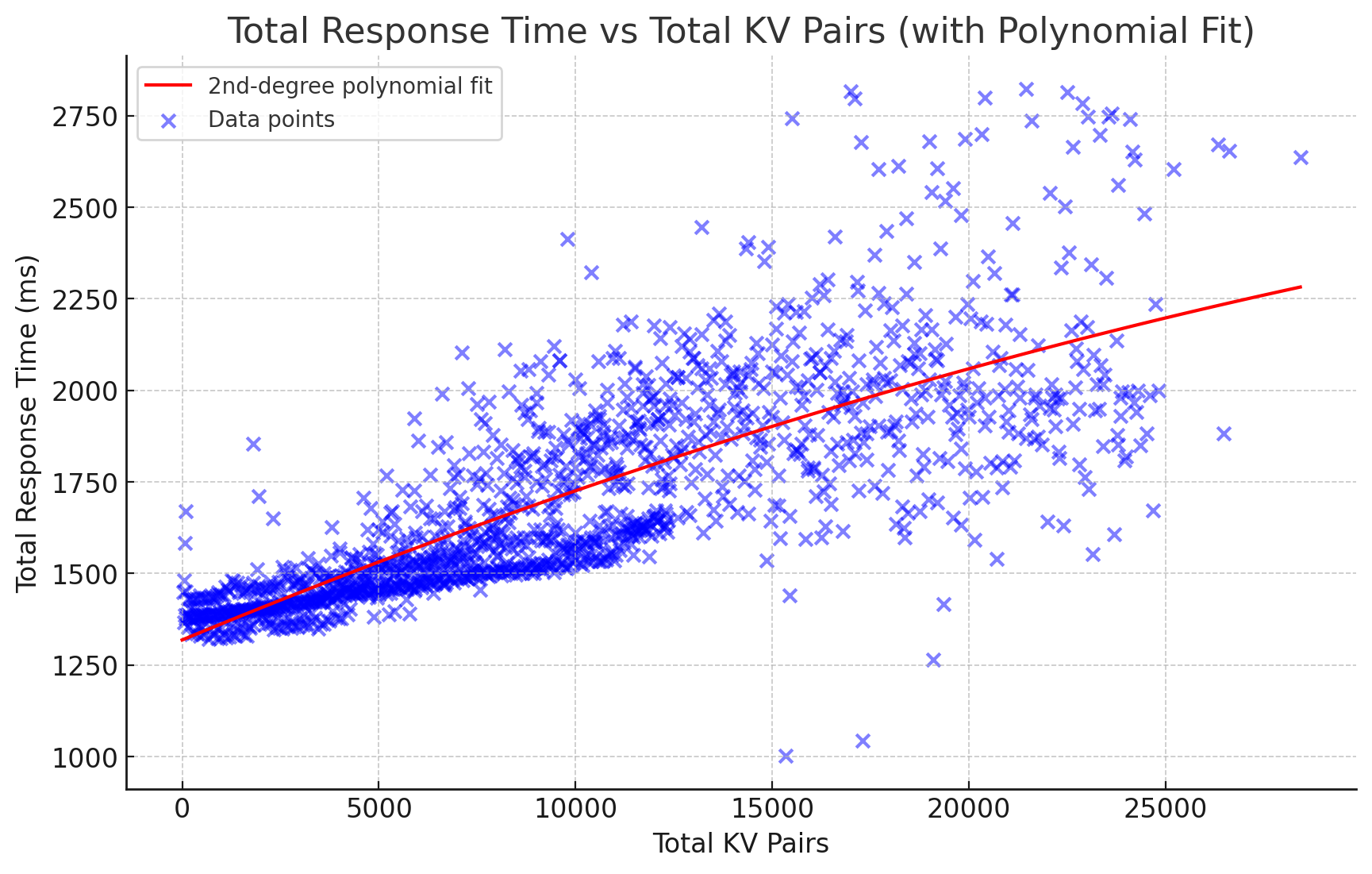}
    \caption{Polynomial Regression on Combined Datasets}
    \label{fig:polyRegression}
\end{figure}

The analysis of the relationship between the total number of key-value pairs (KV pairs) and the total update response time was conducted in a systematic manner using a dataset gathered from the four iterations of our test, each characterized by a different value of  $n$ as explained earlier. The entire analytical process involved several steps, defined as follows:

Exploratory data analysis was carried out to visualize the general trend in the data. Four scatter plots where generated with the total number of KV pairs on the x-axis and the total update response time on the y-axis, each corresponding to different values of  $n$, shown in figure \ref{fig:4datasets}.

The individual datasets corresponding to different cycles were consolidated into a single dataset to facilitate a comprehensive analysis. To enhance the robustness of the subsequent regression analysis, outliers were identified and removed using the Interquartile Range (IQR) method. The bounds for outliers were defined as:
\[
\text{{Lower bound}} = Q1 - 1.5 \times \text{{IQR}}
\]
\[
\text{{Upper bound}} = Q3 + 1.5 \times \text{{IQR}}
\]
where \(Q1\) and \(Q3\) represent the first and third quartiles, respectively, and IQR is given by \(Q3 - Q1\).

Following the data cleansing, a polynomial regression analysis was undertaken to fit a 2nd-degree polynomial curve to the data. The general form of a 2nd-degree polynomial equation is expressed as:
\[
y = ax^2 + bx + c
\]
where \(y\) is the total update response time, \(x\) is the total number of KV pairs, and \(a\), \(b\), and \(c\) are the coefficients determined through the regression analysis.

The fitted polynomial equation to our data is given by:
\[
y = 1.31855648 \times 10^3 + (4.44412352 \times 10^{-2})x - (3.70701991 \times 10^{-7})x^2
\]
Rounding to 2 decimal points:
\[
y = 1.31 \times 10^3 + (4.44 \times 10^{-2})x - (3.77 \times 10^{-7})x^2
\]

\section{Legacy System $R_{ctp}$}
Parallelly, the legacy system's efficiency in adding transactions to a Merkle Trie was tested. Owing to the non-cumulative nature of the legacy system, where a fresh Trie is constructed in each cycle, we hypothesized a logarithmic relationship relative to the number of incoming transactions due to the growth nature of a Merkle Trie. The experiment involved progressively increasing the transaction count in each cycle to examine the correlation between the speed of Trie construction speed and its size. 

\begin{figure}
    \centering
    \includegraphics[width=0.75\linewidth]{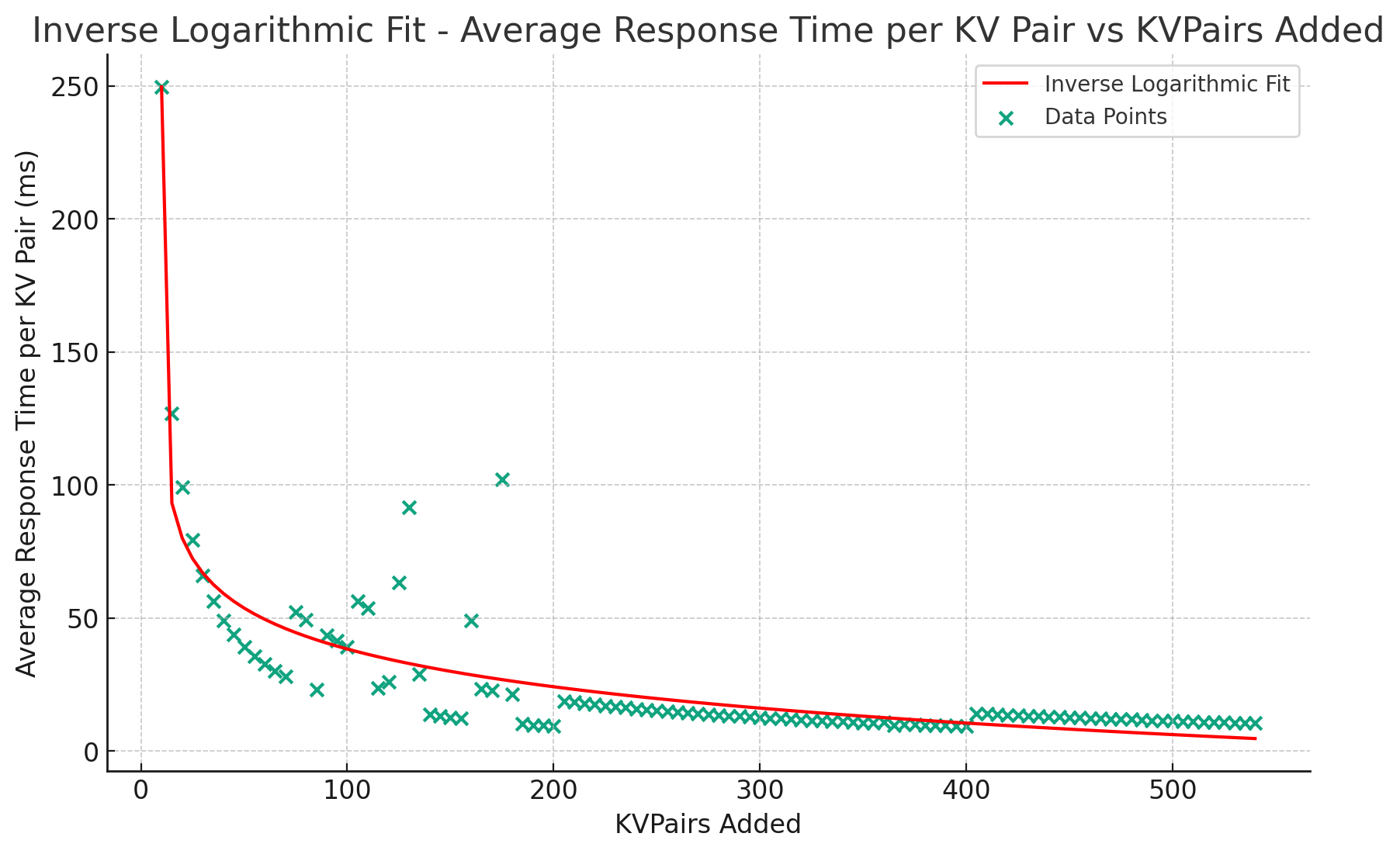}
    \caption{Enter Caption}
    \label{fig:enter-label}
\end{figure}
 
A critical metric of interest is the average response time per key-value (KV) pair addition as a function of the number of KV pairs added per cycle. The data collected over several cycles should present a rich source to understand the underlying dynamics of the response time behaviour in our legacy system.

To elucidate the relationship between the average response time per KV pair and the number of KV pairs added, we initially visualized the data through a scatter plot, which revealed a non-linear trend. Recognizing the complexity of the relationship, we sought to find a function that would accurately depict the trend observed in the data.

The data clearly showed an inverse logarithmic relationship therefore we attempted to fit such a curve to the data. The function is formulated as:

\[
y = a - b \times \ln(x - c)
\]

where \(y\) represents the average response time per KV pair, \(x\) is the number of KV pairs added, and \(a\), \(b\), and \(c\) are parameters to be estimated through curve fitting. The natural logarithm, denoted by \(\ln\), introduces a transformation that can capture diminishing returns, a behaviour suggested by the data.

To find the optimal parameters that minimize the residual error, we employed a non-linear least squares method to fit the function to the data, yielding the result:

\[
y = 123.78 - 18.99 \times \ln(x - 9.99)
\]

The resulting fit presented a curve that adeptly traced the trend in the data, providing a mathematical representation of the relay service's performance as a function of the number of KV pairs added. This function thus offers a valuable tool for predicting the average response time for given inputs, aiding in the optimization and tuning of the relay service's performance.

It is pertinent to note that the function is defined for \(x > c\), which ensures the argument of the logarithm remains positive, averting undefined mathematical operations. This is due to our testing nature starting off at $n$=10 before we iterated through higher numbers of $n$.

\section{Proof of Provenance Retrieval Processing Rate ($R_{pr}$)}
Our enquiry progressed to the investigation of the Proof of Provenance Retrieval Processing rate ($R_{pr}$), a critical determinant in ascertaining the superiority of the novel system over the legacy system. A pivotal assumption in this context is the requisite of the relay service to furnish a proof within the same cycle it was instituted. While this facilitates an implicit null proof for the preceding cycle, the rationale behind this trade-off warrants exploration. Contrarily, the legacy system affords proof retrieval independent of the cycle's processing duration, albeit necessitating multiple null proofs from antecedent cycles.

\subsection{Novel System Evaluation}

For the novel system, the test mirrored the $R_{ctp}$ analysis, with a secondary set of requests dispatched post the receipt of responses to the initial requests. This sequential executor, triggered immediately following the initial responses, adapts dynamically to any delays in the primary responses. The metric of interest here is the average response time, indicative of $R_{pr}$, in relation to the number of KV pairs stored within the Trie, $N$. 

\begin{figure}
    \centering
    \includegraphics[width=0.75\linewidth]{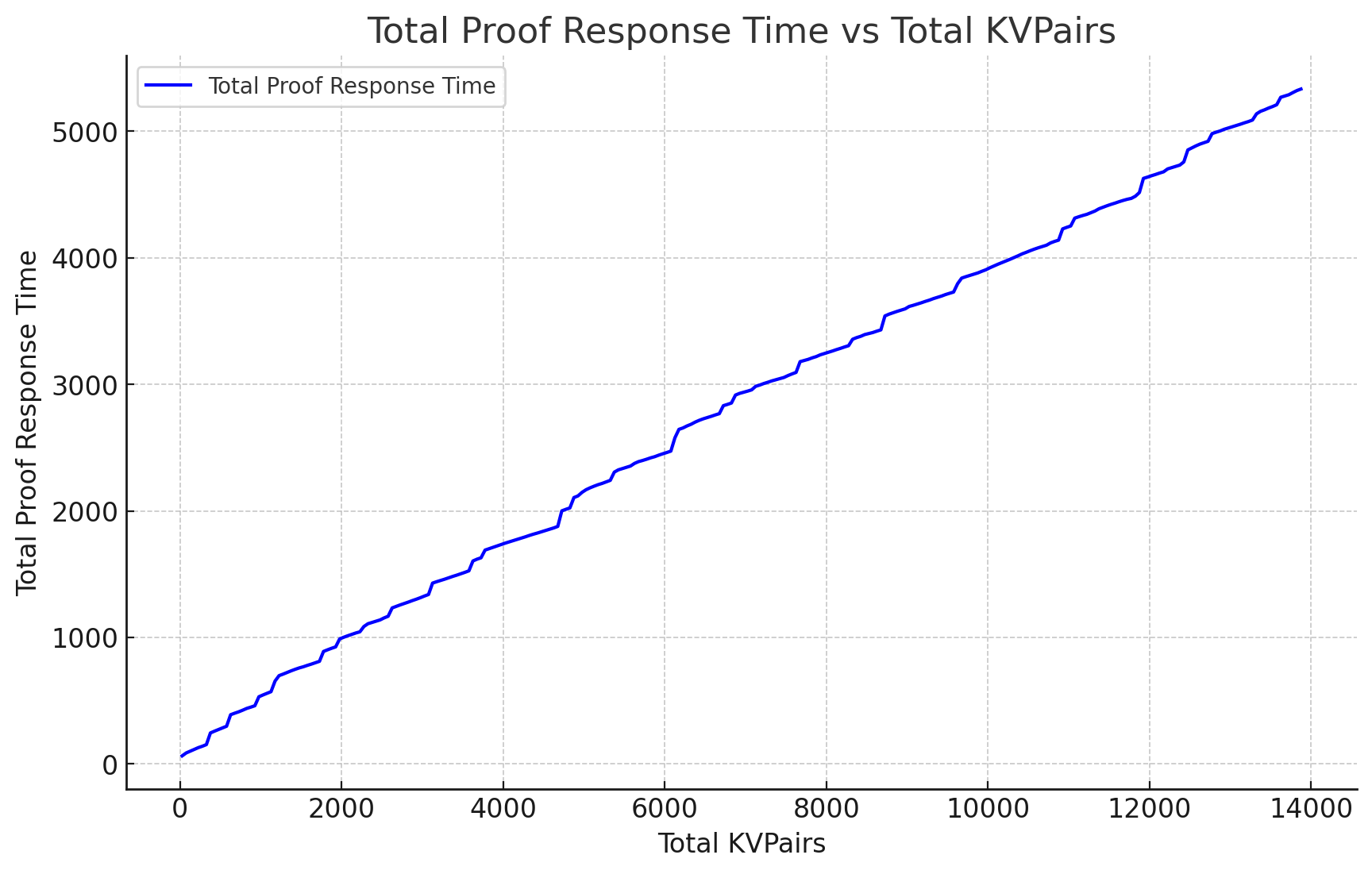}
    \caption{Total Proof Response Time vs Total KV Pairs for Novel Design}
    \label{fig:NovelProofResponse}
\end{figure}
\begin{figure}
    \centering
    \includegraphics[width=0.75\linewidth]{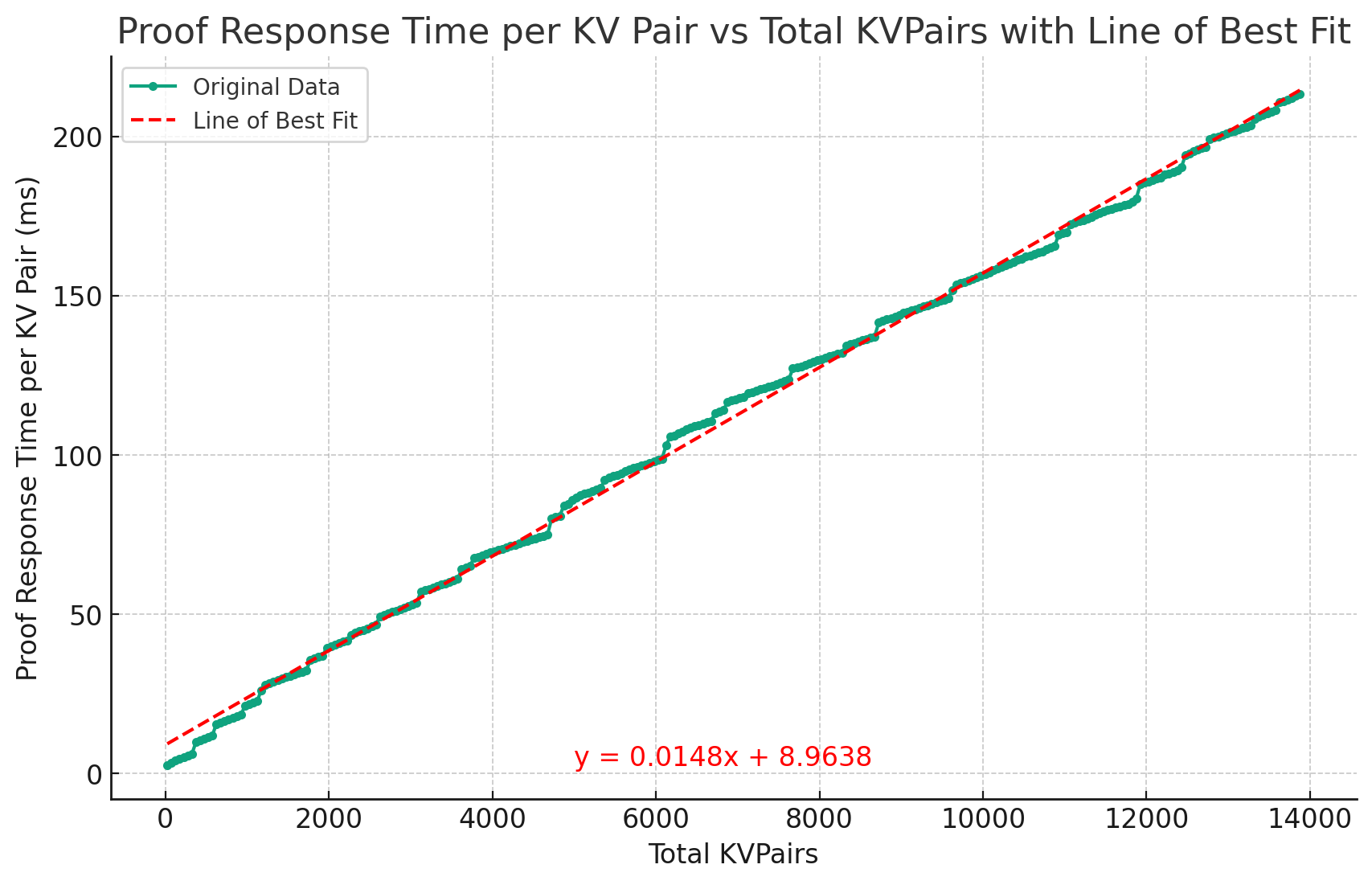}
    \caption{Novel Proof Response Time with a line of best fit for Single KV Pair}
    \label{fig:NovelProofResponseLBF}
\end{figure}

Initially, the dataset was scrutinized to comprehend the underlying structure and the variables it encompassed. We plotted our critical metrics on \ref{fig:NovelProofResponse}, Total Proof Response Time over Total KV Pairs. This gave us an initial understanding of the relationship however did not give us the critical data we need for extrapolation later on. 

To attain a granular understanding of the proof response time with respect to an individual KV pair addition, a new metric was devised. The total proof response time was divided by the number of KV pairs added in within the cycle, yielding a metric representing the proof response time for a single KV pair. 

The newly derived metric was then plotted against the total number of KV pairs stored in the Trie, employing the total KV pairs as the independent variable (x-axis) and the proof response time per KV pair as the dependent variable (y-axis). Each data point on the graph represented a cycle, illustrating the evolution of the proof response time per KV pair over different cycles \ref{fig:NovelProofResponseLBF}.

To capture the underlying trend and to facilitate predictive analysis, a line of best fit was superimposed on the scatter plot. Utilizing a linear regression approach, the line was defined by the equation:

\[
y = 0.015\text{N} + 8.96
\]

where:
\begin{itemize}
    \item \( y \) is the proof response time per KV pair (in milliseconds)
    \item \( N \) is the total number of KV pairs stored in the Trie
\end{itemize}

The final graph encapsulates the analytical process, depicting the proof response time per KV pair as a function of the total number of KV pairs \ref{fig:NovelProofResponseLBF}.

\subsection{Legacy System Evaluation}
Contrastingly, the legacy system demanded a divergent approach to derive $R_{pr}$, revolving around the interceding cycles from the asset's inception to its assimilation in the current cycle's Merkle Trie. The experimentation spanned a thousand cycles to capture a reliable approximation of the $R_{pr}$ relationship. The two-phase testing involved storing the responses from the initial burst of KV pair additions for subsequent utilization in the second phase, where Proofs of Provenance (PoP) were solicited based on a varying delta of cycle count, exploring the relationship of the legacy system to \(R_{pr}^{\text{legacy}}\).

In the exploration of the relationship between the cycle variable and two dependent variables, Total Proof Time and Total Proofs, we initiated the analytical process by plotting scatter graphs to visually inspect the existing relationships.

Post the initial analysis, it was observed that the Total Proof Time data contained outliers which potentially could skew the predictive modeling. To address this, we employed the Interquartile Range (IQR) method to filter outliers, defining them as points lying more than \(1.5 \times \text{IQR}\) away from the first and third quartiles, identical to our $R_{ctp}$ analytical method explained earlier. 

\begin{figure}
    \centering
    \includegraphics[width=0.75\linewidth]{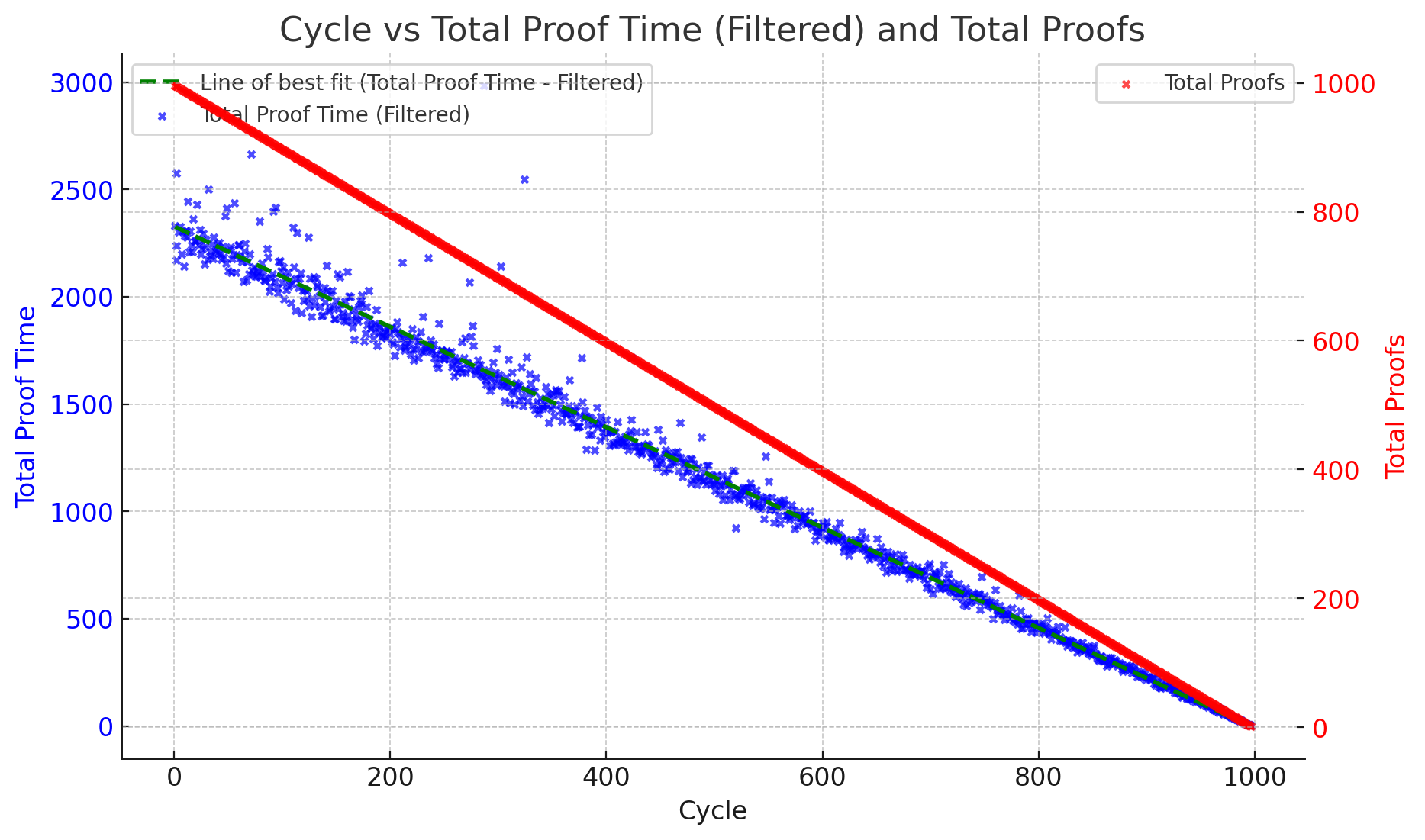}
    \caption{Legacy Systems PoP Retrieval}
    \label{fig:LegacyPOP}
\end{figure}

Subsequently, we introduced a line of best fit to model the relationship between the cycle and the Total Proof Time, shown in figure \ref{fig:LegacyPOP}. The line was derived using a first-degree polynomial fit, mathematically represented as:

\[
y = ax + b
\]

where:
- \( y \) is the predicted Total Proof Time
- \( x \) is the Cycle
- \( a \) and \( b \) are the coefficients derived from the polynomial fit

For our dataset, the coefficients were determined to be approximately \( a = -2.34 \) and \( b = 2328.04 \), resulting in the equation:

\[
y = 2328.04 - 2.34x
\]

Moreover, to concurrently analyze the Total Proofs variable, a dual Y-axis plot was created. This approach maintained the focus on the Total Proof Time while incorporating an additional variable for a comprehensive view.

\section{Discussion}
\label{Chap6}
Our Discussion chapter shall explore the relationships found in our results and analysis sections and subsequently cross reference this with our predicted methodology section. From there we will then hypothesise and extrapolate our mathematical modelling to see if our novel System is truly an enhancement of that of our legacy system. 

\subsection{Analytical Comparison of the Proof Retrieval Systems}

\subsubsection{Legacy System}

The legacy system exhibits a linear dependency of the response time on \( \Delta C \), represented by the equation
\[
R_{pr}^{\text{legacy}} = 2328.04 + 2.34\Delta C,
\]
where \( R_{pr}^{\text{legacy}} \) is the proof response time in milliseconds.

\subsubsection{Novel System}
Contrastingly, the novel system showcases a response time function given by
\[
R_{pr} = 8.96 + 0.015N,
\]
where the relationship between \( N \), \( n \), and \( \Delta C \) is defined as \( N = n \times \Delta C \). Given that \( \Delta C = 1 \) for the novel system, the response time function simplifies to
\[
R_{pr} = 8.96 + 0.015n.
\]
Moreover, the processing time for increasing values of \( N \) is given by 
\[
R_{pr} = 1.31 \times 10^3 + (4.44 \times 10^{-2})N - (3.77 \times 10^{-7})N^2.
\]
\subsection{Derivation of the Proof of Provenance Retrieval Rate Equation}

The Proof of Provenance (PoP) Retrieval Processing Rate ($R_{pr}$) was derived based on the cumulative Merkle Trie model. The initial equation is given as:

\begin{equation}
    R_{pr} = 8.96 + 0.015N
\end{equation}

Where $N$ is the total number of transactions. The total number of transactions $N$ is defined as the sum of incoming transactions over all previous cycles within a period, plus the incoming transactions in the current cycle:

\begin{equation}
    N = n + \sum_{i=1}^{m-1} n_i
\end{equation}

Where $n_i$ represents the incoming transactions in the $i$-th cycle, and $m$ is the number of cycles. Since $n$, the number of transactions accumulated in a single cycle, is influenced by the cycle time ($T_c$) through the relationship $n = \lambda \cdot T_c$, where $\lambda$ is the average rate of incoming transactions, we reformulate the equation for $N$:

\begin{equation}
    N = \lambda \cdot T_c + \sum_{i=1}^{m-1} n_i
\end{equation}

Substituting this into the equation for $R_{pr}$, we get:

\begin{equation}
    R_{pr} = 8.96 + 0.015\left(\lambda \cdot T_c + \sum_{i=1}^{m-1} n_i\right)
\end{equation}

To illustrate the impact of varying transactional throughput ($\lambda$) and period time ($T_p$) on the PoP retrieval rate, a 3D graph was generated. This graph plots the PoP retrieval rate ($R_{pr}$) in seconds against the period time ($T_p$) in hours and the transactional throughput ($\lambda$) ranging from 250 to 1000 transactions per second.For our graphical representation, we made the following assumption:

\begin{itemize}
    \item The cycle time ($T_c$) was fixed at 1 second. This assumption simplifies the model by maintaining a constant interval for transaction accumulation.
\end{itemize}

\subsubsection{Novel System}
The novel system leverages a cumulative build approach, wherein each cycle incorporates the transactions into a continuously building Merkle Trie, inherently restricting \( \Delta C \) to a constant value of 1. This restriction makes the assumption that the CBDC is withdrawn within the same period in which it was transacted. To remove this assumption we must use our existing collected data to map out what is the most optimal parameters we need for the period of time given differing values of \(n\) and then subsequently \(N\), both influenced by the rate of transactions coming into our relay. 

This strategy ensures a consistently low response time, with the only variable being \( n \), the number of KV pairs to be added to the Trie per cycle. However, this advantage comes with the caveat of a more complex processing time, which initially increases with \( N \) but at a diminishing rate due to the quadratic term in the equation inferred from the exponential relationship the number of transactions the Merkle Trie can store in relation to its depth, which is an indicator of the transaction processing speed.

\subsubsection{Legacy System}
In contrast, the legacy system constructs a new Merkle Trie in each cycle, necessitating the retrieval of proofs from all preceding Merkle Tries to validate a proof, thereby incrementally increasing \( \Delta C \) with each cycle. This results in a response time that escalates with the progression of cycles, introducing a potential bottleneck in proof retrieval as the number of cycles grows.

\subsection{Evaluation}
To present the most optimal system we must evaluate our novel system to compare against our traditional architecture in terms of time passed since the withdrawal of CBDC to the transaction of CBDC and see which one has faster proof retrieval times. Our legacy system has a linear relationship with the number of cycles and the proof retrieval time however our novel system is more complex. To better demonstrate this we have calculated differing times of \(T_p\), period time, and its effect on the proof retrieval time. By adding this extra layer of complexitiy and as such removing the restriction of \(\Delta C\) being a constant value we gain a critical insight into the performance of our system over extended periods of time.

The derivation of the proof retrieval time equation for our novel system begins with the foundational assumption of a linear relationship between the retrieval time and the number of Key-Value (KV) pairs processed. This relationship is encapsulated in the following equation:

\[
R_{pr} = 8.96 + 0.015(\frac{\lambda \times T_p}{m}).
\]

Where:
\begin{itemize}
    \item \( R_{pr} \): Proof Retrieval Rate
    \item \( \lambda \): 
    ..
    \item \( T_p \) is the period time, reflecting the duration for which transactions are accumulated.
    \item \( KV_{rate} \) is the transaction rate, indicating the number of KV pairs added to the system per unit of time.
\end{itemize}

This equation assumes a direct proportionality between \( R_{pr} \) and the product of \( T_p \) and \( KV_{rate} \), suggesting that the retrieval time increases linearly with the total number of transactions processed over a given period. The term \( R_{base} \) provides a constant to account for the inherent processing time of the system, independent of the transaction volume. This model presents a comprehensive yet adaptable framework to predict the proof retrieval time, accommodating variations in transaction rate and processing duration.

\begin{figure}
    \centering
    \includegraphics[width=0.5\linewidth]{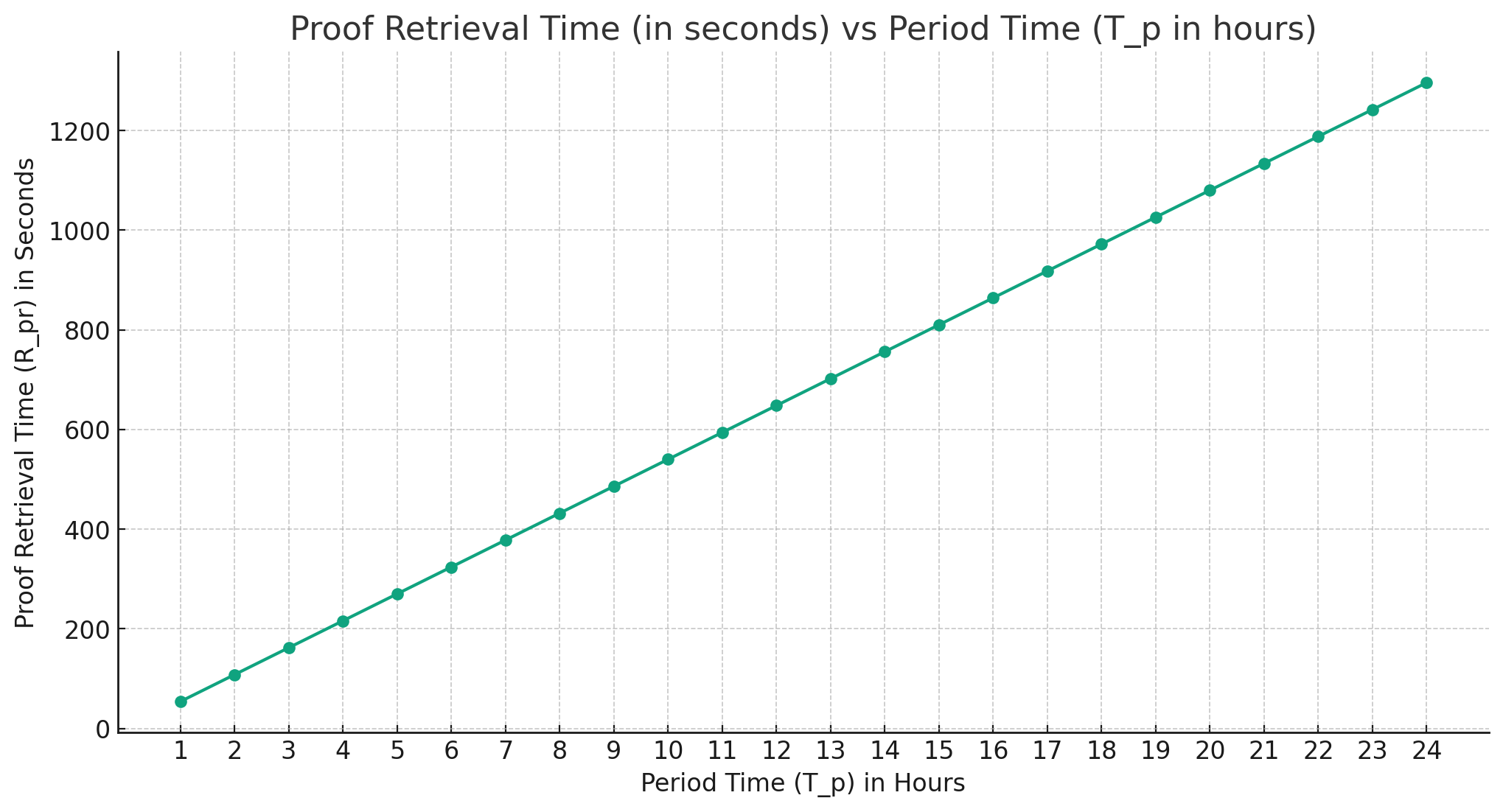}
    \caption{Enter Caption}
    \label{fig:enter-label}
\end{figure}

\section{Limitations}

While the current study offers substantial insights into the Proof Retrieval Time of our newly designed benchmark against the legacy system, it is important to discuss the limitations that could potentially affect the interpretations and generalizations of the findings.

\subsection{Dataset and Extrapolation Concerns}

A primary limitation is inherent in our observation of the \( R_{\text{ctp}} \) values which were derived from varying quantities of \( n \) within our Novel system. Most notably, when \( n=50 \), the data exhibited a decreasing trend towards the end of the dataset. This pattern erroneously suggests a continual decrease in processing time with an increase in the size of \( N \), a phenomenon we know to be false. This brings into question the accuracy of the larger collated dataset, hinting at the possibility of skewed results which, in turn, undermines the reliability of extrapolations beyond the tested ranges.

\subsection{Computational Resources and Environmental Variables}

Our methodology did not initially account for the available computational resources, an oversight stemming from our focused objective to compare latency between the novel and legacy systems. The tests were conducted on a MacBook Air equipped with an M1 chip and 8GB of RAM. Despite the competitive nature of the study theoretically justifying this abstraction, the tests might have been influenced by background processes typical in personal computers, thereby potentially affecting the precision of the results. Although we attempted to mitigate this by filtering the results through the IQR range, the influence of uncontrollable environmental variables remains a considerable limitation.

\subsection{Programming Language and Framework Constraints}

The developmental choice of Java as the programming language, albeit motivated by its pre-runtime compilation facilitating a nearly isolated test environment, introduced its own set of limitations. In particular, the employment of the Spring framework for the continuous deployment of our relay service inadvertently restricted the test parameters due to the inherent constraints of Bean dependencies in managing variable sizes.

Within our novel system, we stored the entire Cumulative Merkle Trie in one of these variables to expedite data retrieval through in-memory storage. However, this approach reached a bottleneck when Java dependencies halted processes as the variable size escalated, imposing a limit on the testable range of \( N \).

\subsection{Network Limitations}
Within our relay service codebase, we manage two threads: one to continuously collect the KV pairs contained within the incoming requests, and the other to process these KV pairs into a Merkle Trie. At the end of each cycle, there is a critical changeover that temporarily halts the collection process as the bucket of accumulated KV pairs is emptied, resulting in a brief pause where incoming requests receive a status 500 error. 

Despite its minor appearance in the context of thousands of requests spanning numerous cycles, this pause represents a limitation by leading to the occasional loss of requests. Future research should address this issue by developing strategies to efficiently handle the changeover without interrupting the collection process, thereby preventing the occurrence of status 500 errors during testing. 

\chapter{Conclusion} \label{Chap6}

Our discussion elucidates the inherent efficiency of the novel system in maintaining a lower response time compared to the legacy system, a benefit derived from its cumulative building strategy that effectively limits \( \Delta C \) to 1 in every cycle. 

However, it is pivotal to note the quadratic nature of the processing time in the novel system, which, while initially increasing with \( N \), does so at a diminishing rate, introducing a nuanced balance between proof retrieval time and processing time. This balance is central to understanding the operational efficiency of the system, potentially offering opportunities for optimization.

Conversely, the legacy system faces a progressively increasing \( \Delta C \), resulting in an escalating response time that could potentially become a bottleneck in the system's proof retrieval efficiency as cycles advance however a more constant processing rate.

The novel system thus stands as a more scalable solution, capable of handling a larger number of transactions more efficiently, while sustaining a lower response time. This renders it a formidable alternative to the legacy system, showcasing a strategic advantage, particularly in environments characterized by high transaction volumes and numerous cycles, such as that for a national retail payments infrastructure.

As such, the novel system unequivocally emerges as a superior solution in sustaining optimal performance in proof retrieval tasks, championing efficiency, and scalability in high-demand scenarios, albeit with a more complex processing time behaviour that warrants careful consideration in the processing power available.

\section{Future Work}

The current study has laid a substantial groundwork in understanding the Proof Retrieval Time of the newly designed system compared to the legacy system. To build upon this research and foster more robust and generalizable findings, future studies should aim to address the identified limitations in the following ways:

\subsection{Enhanced Data Collection and Analysis}

Future work could engage in a more meticulous data collection process to prevent the skewness observed in the current dataset, especially when \( n = 50 \). Utilizing a more diverse range of \( n \) values and potentially leveraging advanced statistical methods for data analysis could aid in achieving a dataset that allows for accurate extrapolations. Moreover, exploring different data preprocessing techniques could enhance the reliability of the findings derived from larger datasets.

\subsection{Resource-Optimized Testing Environment}

To alleviate the concerns arising from the utilization of personal computing resources, future studies should contemplate employing a dedicated testing environment with optimized resources. This would entail conducting tests on systems with dedicated resources to eliminate the influence of background processes and ensure consistency in the results. Additionally, further research could explore the optimization of the testing methodology to mitigate the impacts of environmental variables.

\subsection{Refined Development and Deployment Strategies}

Considering the limitations induced by the choice of programming language and framework, subsequent research could explore alternative development environments and frameworks that offer greater flexibility in managing variable sizes. 

Furthermore, revisiting the strategy for storing the Cumulative Merkle Trie could be beneficial. Future studies might explore efficient data storage and retrieval strategies, possibly considering database solutions that allow for faster data retrieval without overwhelming the system resources. This could potentially involve leveraging more efficient data structures or optimizing the existing ones to accommodate larger values of \( N \) without encountering system halts.

\section{Major Work for the Testing of our DLT System}
As mentioned in the methodology section, we successfully developed a functioning DLT system, fundamentally built on the Raft consensus algorithm and an atomic register. Despite the successful implementation, the system has not undergone extensive testing. This oversight primarily stemmed from the perception that the DLT system does not significantly impact the proof of provenance retrieval time. Specifically, reaching consensus on a single Merkle Root across multiple nodes is considerably faster than constructing a complex Merkle Trie structure. However, this premise warrants a more comprehensive analysis, for which we propose the following framework.

\subsection{Proposed DLT Test}

The pivotal objective of the DLT test is to optimize consensus speed across a substantial number of nodes, aiming to find the equilibrium point where the node consensus speed aligns with the relay service's latency. Achieving this balance would facilitate a framework where each Merkle Root computed in every cycle can be seamlessly integrated into the ledger through our DLT, without causing delays due to consensus time.

We hypothesize that the consensus speed will predominantly be influenced by the node count in the system, asserting that a higher node count augments the system's security but potentially harms the consensus speed of the system. 

It is worth considering that enhancing the node count might necessitate a compromise, excluding some Merkle Roots generated by our novel relay system from being sent to the DLT. This compromise is possible due to the cumulative nature of the Trie, which inherently represents all previously computed roots in the most recent root of the Trie, thereby allowing for the potential to maintain a higher node count without sacrificing the integrity of the information stored in the DLT.

However, this approach is incompatible with our legacy system, prompting a potential reconsideration of the atomic register design. In this redesign, we may envisage a system where, instead of transmitting a single root, multiple roots are dispatched, akin to traditional blockchain designs where blocks encompass numerous transactions. This alteration, while aligning with conventional practices, would entail a trade-off expected to be a surge in consensus times stemming from the augmented data volume, posing possible consideration for future developmental trajectories.
\bibliographystyle{unsrtnat}
\raggedright
\bibliography{references}  






\end{document}